\documentclass[floats,aps,amsfonts,notitlepage,superscriptaddress,twocolumn,eqsecnum,nofootinbib]{revtex4-1}
\usepackage{ifpdf}
\usepackage[utf8]{inputenc}
\usepackage[UKenglish]{babel}
\usepackage{graphicx}
\usepackage[colorlinks]{hyperref}
\usepackage{amsmath, amsthm, amssymb}
\usepackage{multirow}
\usepackage{url}
\usepackage{subfigure}
\usepackage{color}
\usepackage[usenames,dvipsnames,svgnames,table]{xcolor}
\usepackage{tabularx}

\definecolor{CiteColor}{rgb}{0,0.5,0}
\hypersetup{citecolor=CiteColor}
\definecolor{RefColor}{rgb}{0.55,0,0}
\hypersetup{linkcolor=RefColor}
\definecolor{darkgreen}{rgb}{0.2,0.7,0.2}

\newcommand{\diff}[2]  {\frac{d #1}{d #2}}
\newcommand{\sdiff}[2]  {\frac{d^2 #1}{d #2^2}}

\newcommand{\en}{\mathcal{E}}
\newcommand{\ang}{\mathcal{L}}

\newcommand{\self}{\text{self}}

\newcommand{\eff}{\text{eff}}
\newcommand{\ret}{\text{ret}}
\newcommand{\res}{\text{res}}

\newcommand{\out}{\text{out}}
\newcommand{\inn}{\text{in}}
\newcommand{\elm}{\el m}

\renewcommand{\c}{\cos \omega_r \tau}

\newcommand{\el}{\ell}

\newcommand{\be}{\begin{equation}}
\newcommand{\ee}{\end{equation}}
\newcommand{\ba}{\begin{eqnarray}}
\newcommand{\ea}{\end{eqnarray}}

\renewcommand{\c}{\hskip0.0cm,}
\newcommand{\p}{\hskip0.00cm.}

\mathchardef\mhyphen="2D

\def\prd{Phys. Rev. D}

\def\etal{\textit{et al.}}

\begin{document}

\title{Applying the effective-source approach to frequency-domain self-force calculations}
\author{Niels Warburton}
\affiliation{School of Mathematical Sciences and Complex \& Adaptive Systems Laboratory, University College Dublin, Belfield, Dublin 4, Ireland}
\author{Barry Wardell}
\affiliation{Department of Astronomy, Cornell University, Ithaca, NY 14853, USA}
\affiliation{School of Mathematical Sciences and Complex \& Adaptive Systems Laboratory, University College Dublin, Belfield, Dublin 4, Ireland}

\begin{abstract}
The equations of motion of a point particle interacting with its own field are defined in terms of a certain regularized self-field. Two of the leading methods for computing this regularized field are the mode-sum and effective-source approaches. In this work we unite these two distinct regularization schemes by generalizing traditional frequency-domain mode-sum calculations to incorporate effective-source techniques. For a toy scalar-field model we analytically compute an appropriate puncture field from which the regularized residual field can be calculated. To demonstrate the method, we compute the self-force for a scalar particle on a circular orbit in Schwarzschild spacetime. We also demonstrate the relation between the worldtube and window function approaches to localizing the puncture field to the neighborhood of the worldline and show how the method reduces to the well-known mode-sum regularization scheme in a certain limit. This new computational scheme can be applied to cases where traditional mode-sum regularization is inadequate, such as in calculations at second perturbative order.
\end{abstract}	

\date{\today}

\maketitle

\section{Introduction}

With the age of gravitational wave astronomy almost upon us, interest in the general relativistic two-body problem has surged over the past decade. Two body systems are expected to be amongst the brightest sources of gravitational waves. Despite this, waveform templates are still required in order to accurately disentangle the signal from the detector noise. There now exist a wide range of approaches to modeling general relativistic two-body systems, each applicable to a different regime of the problem.

For the case where one of the components of the system is substantially more massive than the other a perturbative treatment can be made, expanding in powers of the mass ratio. Treating the smaller component as a point particle, the first-order-in-the-mass-ratio dissipative dynamics are now well understood \cite{Mino,Drasco-Hughes,Fujita-Hikida-Tagoshi}, and there has been steady progress in understanding conservative corrections to the orbital motion \cite{Barack-Sago-ISCO-shift,Barack-Sago-precession,Sago-Barack-Detweiler,Shah-etal,Shah-etal:Kerr}. Knowledge of the conservative dynamics has proven particularly fruitful, as it has allowed for exchanges and comparisons with other approaches to the two-body problem \cite{Blanchet-etal-PN-SF-comparision:circular,Favata,Damour-EOB-SF,LeTiec_etal-periastron_advance}. 

It has recently been proposed that the range of mass ratios for which black hole perturbation theory is valid may be broad enough to include so-called intermediate-mass-ratio inspirals \cite{LeTiec_etal-periastron_advance}. Excitingly, such systems are candidates for detection in the forthcoming generation of advanced gravitational wave observatories \cite{LIGO_GW_search}. For future space-based detectors, black hole perturbation theory will be key to modeling so-called extreme-mass-ratio inspirals \cite{Gair_etal:review}.

In both cases it is desired to track the phase evolution of the binary to order unity over many hundreds or thousands of orbits \cite{Gair_etal:LISA_event_rates}. To achieve such a high level of accuracy, it is necessary to include second-order-in-the-mass-ratio corrections to the motion \cite{Hinderer-Flanagan}. Beyond improved waveform models, calculation of the second-order corrections will provide another rich seam of information that can be compared with results from other approaches to the two-body problem. Such a calculation is also timely as the theoretical framework required to make a second-order-in-the-mass-ratio calculation has recently been laid \cite{Detweiler:2nd_order,Pound:2nd_order,Gralla:2nd_order}. Before considering how to approach a second-order calculation, it is instructive to review the methods used at first order.

Owing to the point particle model, the first-order-in-the-mass-ratio metric perturbation is divergent at the particle's location. The appropriate regularization scheme to compute the backreaction, or self-force, on the particle was first laid down by Mino, Sasaki and Tanaka \cite{Mino-Sasaki-Tanaka} and Quinn and Wald \cite{Quinn-Wald} and has been elucidated upon by a number of authors since \cite{Detweiler-Whiting,Gralla-Harte-Wald,Gralla-Wald} (see Poisson \etal~for a review \cite{Poisson-review}). The original regularization scheme involved integrating the retarded Green function from past infinity up to, but not including, the current time. This formulation of the regularization procedure is challenging to work with directly, though there has been progress recently \cite{Casals-Dolan-Ottewill-Wardell,Casals-Dolan-Ottewill-Wardell:Schwarzschild}. Most self-force calculations to-date have used reformulations of the original regularization procedure. The particular details of each scheme depend principally on how the metric perturbation is obtained.

One option is to decompose the retarded metric perturbation into spherical-harmonic modes. One then finds that the individual multipole modes of the metric perturbation are finite at the location of the particle. This observation led to the development of the mode-sum regularization scheme, which subtracts an appropriate singular field from the retarded field mode-by-mode \cite{mode-sum-orig}. For perturbations of Schwarzschild spacetime such a decomposition is also advantageous as the individual multipole modes of the metric perturbation decouple from one another. In the axially symmetric Kerr spacetime the multipole modes remain coupled and, though it is possible to proceed with a calculation of this form \cite{Dolan:1+1-Kerr}, researchers have devoted considerable effort into 2+1 and 3+1 decompositions. In these formulations the retarded metric perturbation remains divergent at the particle's location and so, in order to compute the self-force, effective-source approaches were devised \cite{Barack-Golbourn,Barack-Golbourn-Sago,Vega-Detweiler}. In these approaches the singular component of the field is approximated in a neighborhood of the particle. This approximation is then used to construct a sourced field equation, so that far from the particle the retarded field is obtained and nearby the particle a residual field, loosely the retarded field minus the singular field, is directly solved for.

The effective-source approach has been successfully applied to 1+1 \cite{Vega-Detweiler}, 2+1 \cite{Dolan-Wardell-Barack,Dolan-Barack:GSF_m_mode} and 3+1 \cite{Diener_etal:SSF_self_consistent} schemes but as yet has not been applied to frequency domain calculations. The primary reason for this is that at first order in the mass ratio it is not necessary; the metric perturbation is finite at the particle within a frequency domain decomposition\footnote{At first order in the mass ratio the effective-source approach is not required within a 1+1 decomposition either. Vega \& Detweiler \cite{Vega-Detweiler} worked with a 1+1 decomposition as a toy model for developing the effective-source approach.}. At second order in the mass ratio, even within a spherical-harmonic decomposition, the metric perturbation is divergent at the particle and, as such, the mode-sum method cannot be applied. Instead an effective-source approach must be pursued. Furthermore, the effects of the second-order metric perturbation will be very small, being suppressed by two orders of the mass ratio relative to test body effects. This suggests a frequency domain treatment of the problem where one encounters \emph{ordinary differential equations} (ODEs) which are relatively easy to compute numerically to high accuracy. Another reason for considering a frequency-domain approach is that time-domain codes for evolving the Lorenz-gauge gravitational perturbation equations exhibit instabilities in the monopole and dipole modes \cite{Barack-Sago-circular,Dolan-Barack:GSF_m_mode}. Though there has been progress in resolving these issues \cite{Dolan-Barack:GSF_m_mode} there is currently no robust formalism to stably evolve the low multipole modes in the time domain. These three observations are the main motivations for developing a frequency-domain effective-source approach to self-force calculations.

To develop the method, we present in this work a toy scalar-field calculation which provides a clear worked example of our approach in Schwarzschild spacetime. In a second paper we will apply the method to Lorenz gauge gravitational perturbations \cite{Wardell-Warburton-Ottewill}. The format of this paper is as follows. In Sec.~\ref{sec:field_eqs} we present the scalar field equations and their retarded solution. In Sec.~\ref{sec:regularization} we provide a brief overview of the mode-sum and effective-source approaches to regularization. In Sec.~\ref{sec:effective_source_in_FD} we demonstrate how to apply the effective-source approach to frequency domain calculations, presenting some sample results in Sec.~\ref{sec:results}. We then discuss, in Sec.~\ref{sec:mode_sum_and_effective_source}, the relation between the effective-source approach and the mode-sum method, showing how the latter can be derived from the former by taking appropriate limits. We conclude with a few remarks in Sec.~\ref{sec:concluding_remarks}.

Throughout this work we use geometrized units such that the gravitational constant and the speed of light are equal to unity. We denote the black hole's mass by $M$, use metric signature $(-+++)$ and use standard Schwarzschild coordinates $(t,r,\theta,\varphi)$.
\
\section{Scalar field equation and retarded solution for a particle on a circular orbit }\label{sec:field_eqs}
Consider a particle of mass $\mu$ carrying a scalar charge, $q$, moving on a circular geodesic about a Schwarzschild black hole. In this work we shall ignore the effects of the particle's gravitational self-interaction and focus on the self-force arising from the particle's scalar field. Furthermore, we will not consider the backreaction to the particle's motion arising from this force. Instead we seek to calculate the instantaneous force felt by a particle that has spent its entire past history moving along a circular orbit of fixed radius.

Let us denote the particle's worldline by $x_p^\mu(\tau)$ and its four-velocity by $u^\mu = dx^\mu_p/d\tau$ where $\tau$ is the particle's proper time. For a particle on a circular orbit of radius $r=r_0$, $u^r=0$ and, without loss of generality, we shall assume the motion to be in the equatorial plane ($\theta=\pi/2, u^\theta=0$). Defining $f\equiv1-2M/r$, the particle's (specific) energy, $\en_0=-u_t$, and angular momentum, $\ang_0=u_\varphi$, are given by
\begin{equation}
	\en_0 = f_0\left(1-\frac{3M}{r_0}\right)^{-1/2}\c\quad \ang_0 = \frac{r_0\sqrt{M}}{\sqrt{r_0-3M}}\c
\end{equation}
where hereafter a subscript ``0'' denotes a quantity's value at $r=r_0$. The azimuthal frequency with respect to coordinate time is given by
\begin{equation}
	\Omega_\varphi = \frac{d\varphi_p}{dt} = \left(\frac{M}{r_0^3}\right)^{1/2}\c
\end{equation}
and the azimuthal phase accumulates as $\varphi_p(t) = \Omega_\varphi t$.

As we are considering a scalar field theory, there is a wide scope for choosing the field equation \cite{Quinn}. In this work we proscribe that the particle's scalar field, $\Phi$, obeys the field equation
\begin{equation}
  \label{eq:wave}
	\square \Phi \equiv \nabla^\alpha\nabla_\alpha\Phi = -4\pi \rho\c
\end{equation}
where $\nabla$ is the covariant derivative with respect to the background metric and $\rho$ is the particle's scalar charge density. Modeling the particle as a delta function along its worldline gives
\begin{align}
	\rho(t,r,\theta,\varphi) &= q\int \delta^4(x^\mu - x_p^\mu(\tau) )[-g(x)]^{-1/2}\,d\tau	\nonumber \\
					&= \frac{q}{r_0^2 u^t} \delta(r-r_0)\delta(\varphi-\varphi_p)\delta(\theta-\pi/2)\c
\end{align}
where $g=-r^4\sin^2\theta$ is the metric determinant and the $t$ dependence comes through $\varphi_p$. The source can be decomposed into spherical harmonic and frequency modes in the form
\begin{equation}
	\rho(t,r,\theta,\varphi) = \int\sum_{\el=0}^\infty \sum_{m=-\el}^{\el} \hat{\rho}_{\elm}(r) Y_{\elm}(\theta,\varphi)e^{-i\omega t}\,d\omega\c
\end{equation}
where $Y_{\el m}(\theta,\varphi)= \hat{c}_{\elm} P_{\ell}^m (\cos \theta)e^{im\varphi}$ are the standard scalar spherical harmonics with $P_{\ell}^m$ the associated Legendre polynomial and $\hat{c}_{\elm} =\sqrt{\frac{2\ell+1}{4\pi}\frac{(\ell-m)!}{(\ell+m)!}}$. Our choice of normalization for the spherical harmonics gives
\begin{equation}
	\oint Y_{\elm} (\theta,\varphi) Y^*_{\el'm'}(\theta,\varphi) \, d\Omega = \delta_\el^{\el'} \delta_m^{m'}\c
\end{equation}
where $d\Omega = \sin\theta\, d\theta d\varphi$, $\delta_{n_1}^{n_2}$ is the usual Kronecker delta and a $*$ denotes complex conjugation. The periodicity of the orbit implies that each Fourier mode frequency is an integer multiple overtone of the base orbital frequency, i.e., $\omega=n\Omega_\varphi$ for integer $n$. The radial dependence of the source can then be calculated via
\begin{align}
	\hat{\rho}_{\elm}(r) 	&= \frac{1}{T_\varphi}\int^{T_\varphi}_0 \oint \rho\, \hat{c}_{\elm}P_{\el}^m(\cos\theta) e^{i(n-m)\Omega_\varphi t} \,d\Omega dt	\nonumber	\\
												&= \frac{q}{r_0^2 u^t} \hat{c}_{\elm}P_{\el}^m(0)\delta(r-r_0) \delta^n_m\c
\end{align}
where $T_\varphi = 2\pi/\Omega_\varphi$. We thus see that each azimuthal mode contains a single Fourier harmonic of frequency
\begin{align}
	\omega_m = m\Omega_\varphi\p
\end{align}

Knowing the discrete spectrum of the source we can write the field as a sum over spherical harmonic and Fourier modes in the form
\begin{equation}\label{eq:Phi_decomp}
	\Phi(t,r,\theta,\varphi) = \sum_{\el=0}^{\infty}\sum_{m=-\el}^\el \phi_{\elm}(r) Y_{\elm}(\theta,\varphi) e^{-i\omega_m t}\p
\end{equation}
Decomposing the wave operator into spherical harmonic and frequency modes, and dividing by $f$, we define
\begin{equation}
	\square_{\elm} \equiv \sdiff{}{r} + \frac{2(r-M)}{fr^2}\diff{}{r} + \frac{1}{f}\left(\frac{\omega^2}{f} - \frac{\ell(\ell+1)}{r^2}\right)\p
\end{equation}
The field equation for the radial field, $\phi_{\elm}(r)$ is then given by
\begin{equation}\label{eq:radial_eq}
	\square_{\elm} \phi_{\elm} = \kappa_{\elm}\delta(r-r_0)\c \quad \kappa_{\elm} = -\frac{4\pi q}{\en_0 r_0^2} \hat{c}_{\elm} P_{\el}^m(0)\p
\end{equation}
This equation can be solved for mode-by-mode, with the retarded solution constructed by selecting appropriate boundary conditions on the horizon and at spatial infinity. As the source is confined to the equatorial plane, only the modes with $\el+m$ even contribute to the sum in Eq.~\eqref{eq:Phi_decomp}.

Let us denote the inner and outer homogeneous solutions to Eq.~\eqref{eq:radial_eq} by $\tilde{\phi}^-$ and $\tilde{\phi}^+$, respectively. For the radiative modes ($m\neq0$) we introduce the tortoise radial coordinate, $r_*$, defined by $\tfrac{dr_*}{dr}=f^{-1}$, giving explicitly
\begin{equation}
	r_* = r + 2M\log\left(\frac{r}{2M}-1\right)\c
\end{equation}
where we have specified the constant of integration such that $r_*$ and $r$ coincide at $r_*=r=4M$. In this coordinate system spatial infinity is at $r_*=\infty$ and the event horizon is at $r_*=-\infty$. The asymptotic boundary conditions for the retarded field are then given by
\begin{align}\label{eq:BCs_asymptotic}
	\tilde{\phi}_{\elm}(r_*\rightarrow\pm\infty) = e^{\pm i\omega r_*}\c
\end{align}
which ensures radiation is outgoing at spatial infinity and ingoing at the event horizon. In general there are no closed form solutions for the radiative modes, and so we will solve for these modes numerically as we describe below in Sec.~\ref{sec:results}.

The time independence of the static ($m=0$) modes implies that the advanced and retarded solutions are the same with regularity at the boundaries being sufficient to select the correct homogeneous solutions. In this case it is also possible to give analytic solutions:
\begin{align}
	\tilde{\phi}^+_{\el0} 	&= Q_\ell(r/M-1) \label{eq:m=0_out}\c\\
	\tilde{\phi}^-_{\el0} 	&= P_\ell(r/M-1)	\label{eq:m=0_in}\c
\end{align}
where $P_\ell$ and $Q_\ell$ are the Legendre polynomials of the first and second kinds, respectively. 

For both the radiative and static modes the inhomogeneous solution to Eq.~\eqref{eq:radial_eq} can be constructed via the standard variation of parameters approach. With this method the retarded inhomogeneous field, $\phi_{\elm}^\ret$, is constructed via
\begin{equation}\label{eq:phi_inhom_variation}
	\phi_{\elm}^\ret(r) = c_{\elm}^{+\ret}(r) \tilde{\phi}_{\elm}^+(r) + c_{\elm}^{-\ret}(r) \tilde{\phi}_{\elm}^-(r)\p
\end{equation}
The weighting functions, $c_{\elm}^{\pm\ret}(r)$, are given by
\begin{align}
	c_{\elm}^{+\ret}(r) &= \int^r_{2M} \frac{\tilde{\phi}_{\elm}^-(r')}{W(r')} S_{\elm}^\ret\, dr'\c	\\
	c_{\elm}^{-\ret}(r) &= \int^\infty_r \frac{\tilde{\phi}_{\elm}^+(r')}{W(r')} S_{\elm}^\ret\, dr'\c
\end{align}
where the source $S_{\elm}^\ret=\kappa_{\ell m}\delta(r-r_0)$ and $W(r)$ is the Wronskian of homogeneous solutions
\begin{equation}
	W(r) = \tilde{\phi}_{\elm}^-  \tilde{\phi}_{\elm,r}^+ - \tilde{\phi}_{\elm}^+ \tilde{\phi}_{\elm,r}^-\p
\end{equation}
The delta function in the source means the integration can be done analytically and the inhomogeneous solution can be written explicitly as
\begin{align}\label{eq:phi_ret}
	\phi_{\elm}^\ret(r) = \left\{
     \begin{array}{lr}
       \phi_{\elm}^{+\ret}(r) &\quad  r \ge r_0 \c\\
       \phi_{\elm}^{-\ret}(r) &\quad  r \le r_0\c
     \end{array}
   \right.
\end{align}
where
\begin{equation}\label{eq:phi_inhom_matching}
	\phi_{\elm}^{\pm\ret} = c_{\elm0}^{\pm} \tilde{\phi}_{\elm}^\pm \,, \quad \text{with} \quad c_{\elm0}^\pm = \kappa_{\ell m}\frac{\tilde{\phi}^\mp_{\elm0}}{W_0}\,,
\end{equation}
and $W_0$ is the Wronskian evaluated at $r=r_0$.

\section{Regularization}\label{sec:regularization}

An appropriate regularization procedure for calculating the self-force on a point particle coupled to a scalar field in a curved spacetime was first given by Quinn \cite{Quinn}. Later, Detweiler and Whiting gave an alternative perspective to regularization \cite{Detweiler-Whiting}. In their description the self-force is computed from a regular field, $\Phi^R$, via
\begin{equation}\label{eq:Fself_def}
	F^\self_\alpha(x_p) = q \nabla_\alpha \Phi^R(x_p)\p
\end{equation}
The regular field is constructed by subtracting an appropriate singular field, $\Phi^S$, from the usual retarded field $\Phi^\ret$, i.e.,
\begin{equation}\label{eq:PhiR_def}
	\Phi^R(x_p) = \lim_{x\rightarrow x_p}\left[\Phi^\ret(x) - \Phi^S(x)\right]\p
\end{equation}
The precise construction of an appropriate singular field is discussed at length in Refs.~\cite{Poisson-review,Detweiler-Whiting}. One of the key features of the three fields $\Phi^{\ret/R/S}$ is that they obey the field equations
\begin{equation}
	\square \Phi^{\ret/S} = -4\pi\rho\,,\qquad \square \Phi^R = 0\c
\end{equation}
from which we see that both $\Phi^\ret$ and $\Phi^S$ diverge in the same way at the particle's location, whereas their difference, $\Phi^R$, remains finite. From Eqs.~\eqref{eq:Fself_def} and \eqref{eq:PhiR_def} we can write the self-force as
\begin{align}\label{eq:F_self_ret-S}
	F^\self_\alpha(x_p) &= q\lim_{x\rightarrow x_p}\left[\nabla_\alpha(\Phi^\ret(x) - \Phi^S(x))\right]		\nonumber \\
											&= \lim_{x\rightarrow x_p}\left[F^\ret_\alpha(x) - F^S_\alpha\right(x)]\c
\end{align}
where
\begin{align}
	F^{\ret/S}_\alpha(x) \equiv q \nabla_\alpha \Phi^{\ret/S}(x)\p
\end{align}
The divergence of $\Phi^{\ret/S}$ at the particle makes this equation challenging to work with directly and so over the years it has been recast into forms more amenable to practical calculation. Two of these recastings, the mode-sum scheme and effective-source approach, we discuss now.

\subsection{Mode-sum scheme}

The first step in the mode-sum approach to practical regularization is to decompose the right hand side of Eq.~\eqref{eq:F_self_ret-S} into spherical-harmonic modes. The key feature of this decomposition is that the two angular integrals ``smooth out'' the delta-function source so that the individual multipole modes of the field and its derivatives are finite at the particle's location. Explicitly we have
\begin{align}\label{eq:Fself_from_Fl}
	F^\self_\alpha(x_p) = \lim_{x\rightarrow x_p} \sum_{\el=0}^\infty \left[F^{(\ret)\ell}_\alpha(x) - F^{(S)\ell}_\alpha(x)\right]\c
\end{align}
where a superscript $\el$ denotes a quantity's decomposition into spherical-harmonic modes and summed over $m$, i.e.,
\begin{align}
	F^{(\ret/S)\ell}_\alpha = \sum_{m=-\ell}^\ell Y_{\ell m}(\pi/2,\varphi_p)\oint F^{\ret/S}_\alpha Y_{\ell m}^*(\theta,\varphi)\,d\Omega\p
\end{align}
In taking the limit to the worldline in Eq.~\eqref{eq:Fself_from_Fl} some care is required. Though the individual $\ell$-mode contributions $F^{(\ret/S)\ell}_\alpha$ are finite at the particle, in general, their sided limits $r\rightarrow r_0^\pm$ give two different values, which we denote by $F^{(\ret/S)\ell}_{\alpha\pm}$, respectively. For circular orbits there is no closed-form solution for $F^\ret_\alpha$, and typically it is computed numerically. The singular field, on the other hand, is amenable to an analytical treatment. The local structure of the singular field was first analyzed by Mino \etal~\cite{Mino-Sasaki-Tanaka}, and the mode-sum method was developed shortly after by Barack and Ori \cite{mode-sum-orig}. The formula they obtained for the regularized field and self-force is given by
\begin{align}\label{eq:field_mode-sum-formula}
	\Phi^R(x_p) 				&= \sum_{\el=0}^\infty\left(\Phi_\el^\ret - B_\text{field} - C_\text{field}L^{-1}\right)		\c		\\
	F^\self_\alpha(x_p) &= \sum_{\el=0}^\infty\left(F^{(\ret)\ell}_{\alpha\pm} - A_{\alpha\pm}L - B_\alpha - C_\alpha L^{-1} \right)	\c \label{eq:force_mode-sum-formula}	
\end{align}
where $\alpha=\{t,r,\theta,\varphi\}$ and $L=2\el+1$. The $\el$-independent $A,B,C$ are known as regularization parameters and their value is known for generic geodesic orbits in Schwarzschild \cite{mode-sum-orig} and Kerr spacetime \cite{Barack-Ori}. In general $C_\alpha=C_\text{field}=0$ and, for circular orbits, the other nonzero regularization parameters are given by
\begin{align}
	A_{\pm r} 			&= \mp\frac{q^2}{2 r_0^2}\frac{\en_0}{f_0 V}		\c	\\
	B_r							&= \frac{q^2}{r_0^2}\frac{\en^2_0[ E(w)-2 K(w)]}{\pi f_0 V^{3/2}}	\c	\\
	B_\text{field} 	&= q\frac{2K(w)}{\pi\sqrt{\ang_0^2+r_0^2}}\c
\end{align}
where $K(w)=\int^{\pi/2}_0(1-w \sin^2\theta)^{-1/2}\,d\theta$ and $E(w)=\int^{\pi/2}_0(1-w \sin^2\theta)^{1/2}\,d\theta$ are the complete elliptic integrals of the first and second kinds, respectively, and
\begin{align}
	w \equiv \frac{\ang_0^2}{\ang_0^2+r_0^2}\c \quad V \equiv 1+\frac{\ang_0^2}{r_0^2}\p
\end{align}
As the series in Eqs.~\eqref{eq:field_mode-sum-formula} and \eqref{eq:force_mode-sum-formula} is truncated at $\el^{-1}$, the residual series converges like $\el^{-2}$. It is possible to derive higher-order regularization parameters that serve to increase the convergence rate of the mode-sum. It is common practice in mode-sum calculations to fit for these terms numerically. Analytically, Detweiler \etal~\cite{Detweiler-Messaritaki-Whiting} made the first calculation of a higher-order parameter for a scalar particle on a circular orbit about a Schwarzschild black hole, and recently Heffernan \etal~computed the next 3 or 4 parameters for scalar, electromagnetic and gravitational particles for generic geodesic motion in Schwarzschild \cite{Heffernan-Ottewill-Wardell} and Kerr geometry \cite{Heffernan-Ottewill-Wardell:Kerr}.

\subsection{Effective-source approach}

The effective-source approach provides an alternative method for handling
the divergence of the retarded field. Rather than first computing the retarded field and then
subtracting the singular piece as a postprocessing step, one can instead work directly with an
equation for the regular field. This idea, independently proposed by Barack and Golbourn
\cite{Barack-Golbourn} and by Vega and Detweiler \cite{Vega-Detweiler} has the distinct advantage
of involving only regular quantities, making it applicable in a wider variety of scenarios than
the mode-sum scheme.

Using Eq.~\eqref{eq:PhiR_def} to rewrite $\Phi^{\rm ret}$ in terms of $\Phi^R$ and $\Phi^S$, we can
rewrite Eq.~\eqref{eq:wave} as
\begin{align}
	\square\Phi^R &= \square(\Phi^\ret - \Phi^S)	\nonumber \\
								&= -4\pi\rho - \square \Phi^S.
\end{align}
If $\Phi^S$ is exactly the Detweiler-Whiting singular field, then the two terms on the right hand
side of this equation cancel and $\Phi^R$ would be a homogeneous solution of the wave equation.
However, one typically does not have access to an exact expression for $\Phi^S$. Indeed, the
Detweiler-Whiting singular field is defined through a Hadamard parametrix which is not
even defined globally. Instead, the best one can typically do is a local expansion which is valid only in the
vicinity of the worldline. Denoting an approximation to $\Phi^S$ by $\Phi^P$, the corresponding
approximate regular field is a solution of the sourced wave equation with an
\emph{effective source} given by
\begin{equation}
  S_{\rm eff} = -4\pi\rho - \square \Phi^P.
\end{equation}
This effective source is finite everywhere, but has limited differentiability on the worldline.

An additional level of complexity arises from the fact that the approximation to the singular field
is valid only in the vicinity of the worldline. To avoid ambiguities in its definition far from the
worldline, one must ensure that the singular field goes to zero there. This is most easily achieved by
multiplying $\Phi^P$ by a window function, $\mathcal{W}$, with properties such that multiplying it by $\Phi^P$ only modifies terms higher order in the local expansion about the worldline than those which are explicitly given in $\Phi^P$. In our particular case, it suffices to choose $\mathcal{W}$ such that
$\mathcal{W}(x_p) = 1$, $\mathcal{W'}(x_p) = 0$, $\mathcal{W''}(x_p) = 0$ and $\mathcal{W} = 0$ far
away from the worldline. The residual field then obeys
\begin{equation}\label{eq:box_phi_res}
	\square \Phi^\res = - 4\pi\rho - \square(\mathcal{W}\Phi^P) \equiv S_\eff\c
\end{equation}
and has the properties
\begin{gather}
	\Phi^\res(x_p) 	= \Phi^R(x_p), \quad \nabla_\alpha \Phi^\res(x_p) 	= \nabla_\alpha \Phi^R(x_p)\c
\nonumber \\
	\Phi^\res(x) 		= \Phi^\ret(x) \quad \text{for} \quad x \not\in \operatorname{supp}(\mathcal{W})\p
\end{gather}
As the residual field coincides with the retarded field far from the particle we can use the usual retarded field boundary conditions when solving Eq.~\eqref{eq:box_phi_res}.

\section{Effective source in the frequency domain}\label{sec:effective_source_in_FD}

We now describe the application of the effective-source approach to frequency-domain calculations of the self-force. We first compute a suitable puncture. We then show how to compute the effective source using a worldtube-like window function. Finally we demonstrate how the effective-source method can be used to construct the residual field.

\subsection{Construction of the puncture field}

We seek a suitable radial puncture function $\phi^P_{\ell m}$ which must satisfy the property that
when summed over spherical harmonic $\ell, m$ modes it agrees with the Detweiler-Whiting singular
field and its first derivative when evaluated on the worldline. To construct an appropriate
puncture function, we begin with a local coordinate series approximation to the Detweiler-Whiting
singular field, obtained in terms of Riemann normal coordinates centered on the particle's
worldline using the methods of Ref.~\cite{Heffernan-Ottewill-Wardell}. We then adapt methods
developed for the mode-sum regularization approach
\cite{Barack-Ori,Detweiler-scalar-circular,Haas-Poisson-mode-sum,Heffernan-Ottewill-Wardell} to
analytically decompose the result into spherical-harmonic modes. Finally, we transform the
spherical-harmonic components to components in the unrotated coordinate frame.

\subsubsection{Riemann normal coordinate expansion of the Detweiler-Whiting singular field}
Using $(r,\theta',\varphi')$ to represent the spherical
coordinates in a rotated coordinate system, we orient the coordinates such that the particle is
instantaneously located at $(r_0,0,0)$ and the tangent to its worldline is pointing along the
$\hat{\theta}'$ direction. Using $(r,\theta,\varphi)$ to represent the standard spherical
coordinates where the worldline is on the equator, $(r_0,\pi/2,\varphi_p)$, the two coordinate
systems are related by
\begin{eqnarray}
\sin \theta \cos (\varphi-\varphi_p) &=& \cos \theta', \nonumber \\
\sin \theta \sin (\varphi-\varphi_p) &=& \sin \theta' \cos \varphi', \nonumber \\
\cos \theta &=& \sin \theta' \sin \varphi'.
\end{eqnarray}
This is equivalent to a rotation by Euler angles $(\varphi_p, \pi/2, \pi/2)$. The first rotation
(by an angle $\varphi_p$ around the $z$-axis) ensures that the particle is at $\varphi=0$, the
second rotation (by $\pi/2$ about the new $y$-axis) aligns the north pole with the particle's
position, and the third rotation (by $\pi/2$ about the new $z$-axis) orients the coordinates such
that the tangent to the worldline is pointing along $\hat{\theta}'$.

Working in these rotated coordinates, a coordinate series approximation to the Detweiler-Whiting
singular field may generally be written in a way such that all terms have the form $A^{(n+m)}_{i_1
\cdots i_m} \rho_0^n \Delta x^{i_1} \cdots \Delta x^{i_m}$, where $A^{(n+m)}_{i_1 \cdots i_m}$ is a
function of the worldline parameters only, $m+n$ is the order of the term in the expansion, $n$ is
odd,
\begin{equation}
  \label{eq:rho0}
  \rho_0^2 \equiv (g_{ij} + u_i u_j) \Delta x^i \Delta x^j = B(\delta^2 + 1 - \cos \theta'),
\end{equation}
$\Delta x^{i} \equiv (r-r_0, 2 \sin \frac{\theta'}{2} \cos \varphi', 2 \sin \frac{\theta'}{2} \sin \varphi')$, $B$ is a function of the particle's position, four-velocity and $\sin^2 \varphi'$, and
$\delta^2 \propto \Delta r^2$ with the proportionality constant depending on $B$ and $\sin^2 \varphi'$.

For our frequency-domain effective source, it suffices to keep only the leading two orders in the
expansion so that it has the form
\begin{equation}
\label{eq:PhiP}
\Phi^P = q\left(\frac{1}{\rho_0} + \frac{1}{\rho_0^3} A_{ijk}\Delta x^{i} \Delta x^{j} \Delta x^{k} \right).
\end{equation}
This approximation is sufficient for directly computing the regularized self-force without any
postprocessing regularization step. The effect of neglecting the higher-order terms in the
expansion is to limit the mode-sum to quadratic --- as opposed to exponential --- convergence. While
it would be possible to include the higher-order terms in the puncture to improve convergence, it
turns out not to be necessary to do so. As we show later, a better solution in terms of
computational efficiency is to use the standard mode-sum regularization parameters given in
Ref.~\cite{Heffernan-Ottewill-Wardell}.

\subsubsection{Spherical-harmonic decomposition}
We next seek to decompose this puncture into spherical-harmonic modes labeled by $\ell$ and $m'$.
This proceeds in the same spirit as the standard approach to the computation of regularization
parameters \cite{Heffernan-Ottewill-Wardell}. The advantage of working in a rotated coordinate
frame is now apparent; only the $m'=0$ spherical-harmonic modes are nonvanishing on the polar
axis\footnote{Strictly speaking, since the effective source involves a second-order differential
operator acting on $\phi^P_{\ell m}$ one would also require the modes $m' = \pm 1, \pm 2$. However,
that turns out not to be the case here; the spatial portion of the wave operator depends only on
$\ell$, and the time derivatives are given analytically by multiplication by $-i \omega_m$.}
so we can obtain an approximate decomposition valid near the worldline by considering only $m'=0$.
We therefore need to compute the integrals
\begin{align}
\phi^P_{l,m'=0}(t,r) = \int_{-\pi}^{\pi} \int_{0}^{\pi}
  \Phi^P(t, r, \theta', \varphi') Y^*_{l,m'=0} (\theta', \varphi') d\Omega \nonumber \\
 = \sqrt{\frac{2\ell+1}{4\pi}} \int_{-\pi}^{\pi} \int_{0}^{\pi} \Phi^P(t, r, \theta', \varphi') P_{\ell}(\cos\theta') d\Omega\p
\end{align}

The integrals over $\theta'$ are most easily computed by finding expansions of odd-integer powers of
$\rho_0$ in terms of Legendre polynomials \cite{Detweiler-Messaritaki-Whiting}. Using the
generating function for the Legendre polynomials we can expand the right hand side of
Eq.~\eqref{eq:rho0} to get
\begin{eqnarray}
\rho_0^n &= B^{n/2} \sum_{\ell=0}^\infty \mathcal{A}_{\ell}^{{n}/{2}} (\delta) P_{\ell} \left( \cos \theta' \right).
\end{eqnarray}
The integration is then trivially given by the orthogonality relations
\begin{equation}
  \int_{-1}^{1} P_{\ell} \left( x \right) P_{\ell'} \left( x \right) dx = \frac{2}{2\ell+1} \delta_{\ell \ell'}.
\end{equation}
In the traditional mode-sum regularization approach, at this point one would evaluate this at
$\delta = 0$. However, since we require an extended puncture function in a neighborhood of the
worldline we instead use a power series in $\delta$. For the second-order puncture, the required
expansions are
\begin{align}
  \mathcal{A}_{\ell}^{-1/2}(\delta) =& \sqrt{2} - (2\ell+1)\delta + \mathcal{O}(\delta^2)\c \nonumber \\
  \mathcal{A}_{\ell}^{-3/2}(\delta) =& \frac{2\ell+1}{\delta} + \mathcal{O}(1) .
\end{align}

The integration over the azimuthal angle, $\varphi'$, now reduces to integrals which either vanish or
contain only integer and half-integer powers of $\chi \equiv 1 - k^2 \sin^2 \varphi'$, with $k$
being a function of the particle's position and four-velocity. These integrals all yield hypergeometric functions
\begin{equation}
  \frac{1}{2 \pi}\int_0^{2\pi} \chi^{-n} d\varphi' ={}_2 F_1 (n, \frac{1}{2}; 1; k) \equiv \mathcal{F}_n(k).
\end{equation}
For integer powers, these may be trivially evaluated as polynomials in $k$ and $(1-k)^{-1/2}$. For
$n = \pm \frac12$, the integrals are complete elliptic integrals of the first and second kinds,
respectively. For all other half-integer $n$, the hypergeometric functions can be manipulated to
the complete elliptic integral form using the recurrence relation in Eq.~(15.2.10) of
\cite{Abramowitz:Stegun},
\begin{equation}
\mathcal{F}_{p+1} (k) = \frac{p-1}{p \left(k - 1\right)} \mathcal{F}_{p-1}(k) + \frac{1 - 2p + \left(p - \frac{1}{2}\right) k}{p \left(k - 1\right)} \mathcal{F}_p(k).
\end{equation}
The final result is an expression for $\phi^P_{l,m'=0}(t,r)$ as a power series in $\Delta r$ with
coefficients which depend on the worldline both explicitly and through complete elliptic integrals
whose arguments are functions of the worldline.

\subsubsection{Rotation of the coordinate system}
The final step in the construction of the puncture is the conversion of the spherical-harmonic
components in the rotated coordinate frame to components in the unrotated frame. Under a rotation of the coordinate system which is represented by the Euler angles $\alpha, \beta,
\gamma$, the spherical-harmonic components transform according to
\begin{equation}
  \phi_{\ell m} = \sum_{m'=-\ell}^{\ell} D_{mm'}^{\ell} (\alpha, \beta, \gamma) \phi_{\ell m'},
\end{equation}
where $D_{mm'}^{\ell} (\alpha, \beta, \gamma)$ is the Wigner-D matrix \cite{Wigner}. Here, we use
the convention that the Euler angles correspond to a $z-y-z$ counterclockwise rotation and our
convention\footnote{This convention is different from that of 
\emph{Mathematica} \cite{Mathematica} and Wigner \cite{Wigner}. Our $D_{mm'}^{\ell} (\alpha, \beta, \gamma)$ is related to theirs by a change in the signs of $m$ and $m'$ \cite{Rose}.} for $D_{mm'}^{\ell} (\alpha, \beta, \gamma)$ is
consistent with Rose \cite{Rose}. Using these conventions,
the Wigner-D matrix satisfies
\begin{equation}
  D_{m_1 m_2}^{\ell}(\alpha, \beta, \gamma) = e^{-i m_1 \alpha - i m_2 \gamma} D_{m_1 m_2}^{\ell} (0, \beta, 0).
\end{equation}
Since we are including only the $m'=0$ modes, we require the Wigner-D matrix only with $m'=0$,
in which cases it is most conveniently written in terms of the spherical harmonics.
\begin{equation}
  D_{m 0}^{\ell}(\alpha, \beta, \gamma) = \sqrt{\frac{4\pi}{2\ell+1}} Y_{\ell m}^* (\beta, \alpha).
\end{equation}
We are interested in the particular rotation by the angles $(\varphi_0,\pi/2, \pi/2)$ in which
case things simplify even further and the nonconstant piece of the rotation is a trivial phase
factor,
\begin{equation}
  D_{m 0}^{\ell}(\varphi_0, \pi/2, \pi/2) = \sqrt{\frac{4\pi}{2\ell+1}} Y_{\ell m} (\pi/2, 0) e^{-i m \varphi_0}.
\end{equation}

\subsubsection{Puncture for circular orbits in Schwarzschild spacetime}

For the case of a particle in a circular orbit in Schwarzschild spacetime, the local expansion of
the three-dimensional puncture has the simple expression
\begin{align}
\label{eq:PhiP-circ}
\Phi^P =& q\Bigg\{\frac{1}{\rho_0}
 + \frac{1}{\rho_0^3} \Bigg[
   \frac{M}{2(r_0 - 2M)^2} \Delta r^3 \nonumber \\
& \quad + \left(2 r_0- \frac{4 r_0 (r_0 - 2M)}{r_0 - 3M} \chi\right) \Delta r \sin^2 \frac{\theta'}{2}
  \Bigg]\Bigg\},
\end{align}
where
\begin{equation}
\rho_0^2 = \frac{r_0}{r_0-2M} \Delta r^2 + \frac{2r_0^2(r_0-2M)}{r_0-3M} \chi (1-\cos\theta')\c
\end{equation}
and $\chi \equiv 1 - \frac{M}{r_0-2M}\sin^2 \varphi'$. We then have that
\begin{equation}
  \delta^2 = \frac{(r_0-3M) \Delta r^2}{2 r_0 (r_0-2M)^2 \chi}\c
\end{equation}
and
\begin{equation}
  B = \frac{r_0 \Delta r^2}{(r_0 - 2M)\delta^2}.
\end{equation}

Decomposing this puncture into spherical harmonics, we obtain the approximate spherical-harmonic modes of the puncture in the rotated frame as
\begin{align}
	\phi^P_{\el,m'=0} =& q \sqrt{\frac{4\pi}{2\el+1}}\Bigg\{-\frac{(2 \el+1) |\Delta r|}{2 r_0 (r_0-2M)} \sqrt{1-\frac{3M}{r_0}} \nonumber \\
  & +\frac{1}{\pi r_0}\sqrt{\frac{r_0-3 M}{r_0-2 M}} \left[2K
    +\frac{\left(E-2 K\right)}{r_0} \Delta r \right]\Bigg\},
\end{align}
where, recall, $K$ and $E$ are the complete elliptic
integrals of the first and second kinds, respectively, with argument $M/(r_0-2M)$.

To relate this to the puncture in regular Schwarzschild coordinates where the particle is on the
equatorial plane, we simply multiply by the Wigner-D matrix. We then find that the rotation gives the
puncture for the $m\neq0$ modes as a simple rescaling of the $m=0$ mode:
\begin{equation}
	\phi^P_{\elm}(t,r) = \left(\sqrt{\frac{4\pi}{2\el+1}} Y_{\elm}(\pi/2,0) \right) e^{i m \Omega_\varphi t} \phi^P_{\el,m'=0},
\end{equation}
where there is no sum over $\ell,m$. Finally, the explicit dependence on $t$ here makes a further
decomposition into Fourier modes trivial:
\begin{align}
	\phi^P_{\elm}(r) &= \frac{1}{2\pi} \int_{-\infty}^{\infty} \phi^P_{\elm} (t,r) e^{-i \omega t} dt \nonumber \\
   &= \delta(\omega - m \Omega_\varphi) \phi^P_{\elm}(0,r).
\end{align}

\subsection{Construction of the effective source}\label{sec:scalar_eff_source}

There is considerable freedom in the choice of window function, $\mathcal{W}$, used to confine
the definition of the puncture to the neighborhood of the worldline. The most straightforward
option is to use a simple Gaussian window function. This
choice is simple to implement and allows for a smooth transition from the residual field near the
particle to the retarded field far away. The downside is that it leads to a noncompact effective
source, which formally leads to an integral over an infinite domain when using the method of
variation of parameters to construct the inhomogeneous fields. One alternative
would be to use the window function of Vega \etal~\cite{Vega_etal:SF_in_3+1} which leads to a
compact effective source and allows for a smooth transition from the residual to the retarded
field. Another alternative, which also provides a compact source, is to use the worldtube approach
of Barack and Golbourn \cite{Barack-Golbourn}. In this work we opt to use the latter technique but
we approach it from a window function point of view. In doing so we concretely identify the
equivalence of the window function and worldtube approaches.

Making use of the Heaviside $\Pi$ function defined by
\begin{equation}
	\Pi(x) = \left\{
     \begin{array}{lr}
       1 & \quad |x| < 1/2\c\\
       0 & \quad |x| > 1/2\c
     \end{array}
   \right.
\end{equation}
we define our window function to extend from $r=r_a<r_0$ to $r=r_b>r_0$,
\begin{equation}\label{eq:window_func}
	\mathcal{W} = \Pi(x(r)) \quad \text{where} \quad x(r) = \frac{r - (r_b+r_a)/2}{r_b-r_a}.
\end{equation}
The effective source is then given by
\begin{align}\label{eq:Seff_SeffI_SeffB}
	S^\eff_{\elm} =&  \kappa_{\elm} \delta(r-r_0) -\square_{\elm} (\mathcal{W}\phi^P_{\elm}) \nonumber \\
  \equiv& S_{\elm}^I \Pi(x) + S_{\elm}^B\c
\end{align}
where we have separated the effective source into two terms, one coming from the interior of the
puncture region and the other from the boundary. In Schwarzschild spacetime, these two terms are
given explicitly by
\begin{widetext}
\begin{align}
S_{\elm}^I & =  \kappa_{\elm} \delta(r-r_0) -\sdiff{\phi^P_{\elm}}{r} - \frac{2(r-M)}{fr^2}\diff{\phi^P_{\elm}}{r} - \frac{1}{f}\left(\frac{\omega_m^2}{f} - \frac{\ell(\ell+1)}{r^2}\right) \phi^P_{\elm}	 =  \kappa_{\elm} \delta(r-r_0) -\square_{\elm} \phi^P_{\elm}\c			\\
	S_{\elm}^B 	& = -\left[\frac{\delta' \left(x_a\right)+\delta' \left(-x_b\right) }{(r_b-r_a)^2} +\frac{2(r-M) \left(\delta \left(x_a\right)-\delta \left(x_b\right)\right)}{f r^2 (r_b-r_a)} \right]\phi^P_{\elm} - \frac{2\left(\delta \left(x_a\right)-\delta \left(x_b\right)\right)}{r_b-r_a} \diff{\phi^P_{\elm}}{r}\c
\end{align}
\end{widetext}
where
\begin{equation}
	x_a = \frac{r_a-r}{r_a-r_b} \c \qquad x_b = \frac{r_b-r}{r_a-r_b}.
\end{equation}
and a prime denotes differentiation with respect to the argument. Note that the term involving
$\delta(r-r_0)$ can be ignored in practice as it exactly cancels an equal term contained inside
$\square_{\elm} \phi^P_{\elm}$.

\subsection{Construction of the residual field}

In constructing the inhomogeneous residual field we once again turn to the variations of parameters approach. Using the two homogeneous fields as a basis, the inhomogeneous solution is constructed via
\begin{equation}\label{eq:phi_inhom_variation_res}
	\phi_{\elm}^\res(r) = c_{\elm}^{+\res}(r) \tilde{\phi}_{\elm}^+(r) + c_{\elm}^{-\res}(r) \tilde{\phi}_{\elm}^-(r)\c
\end{equation}
where recall $\tilde{\phi}_{\elm}^-$ and $\tilde{\phi}_{\elm}^+$ are the inner and outer homogeneous solutions to the radial equation \eqref{eq:radial_eq}. The weighting functions, $c_{\elm}^{\pm\res}(r)$, are given by
\begin{align}
	c_{\elm}^{+\res}(r) &= \int^r_{2M} \frac{\tilde{\phi}_{\elm}^-(r')}{W(r')} S_{\elm}^\eff\, dr' 		\c	\label{eq:cres+}\\	
	c_{\elm}^{-\res}(r) &= \int^\infty_r \frac{\tilde{\phi}_{\elm}^+(r')}{W(r')} S_{\elm}^\eff\, dr'	\c	\label{eq:cres-}
\end{align}
where $W(r)$ is the Wronskian of homogeneous solutions. In general we only have access to $\tilde{\phi}_{\elm}^\pm$ numerically. The $\delta$ and $\delta'$'s that appear in $S_{\elm}^B$ make it challenging to numerically evaluate the weighting functions $c_{\elm}^{\pm\res}$. Instead, we analytically determine the contributions from the $\delta$ and $\delta'$'s, as functionals of $\tilde{\phi}_{\elm}^\pm$, which leaves the remaining contribution free of singularities and accessible to standard numerical integration routines. Splitting the effective source into boundary and interior terms, as in Eq.~\eqref{eq:Seff_SeffI_SeffB}, and integrating the term containing $S_{\elm}^B$ against some test function, $g(r)$, gives a practical formula for computing the weighting functions:
\begin{align}\label{eq:c_res_out}
	c_{\elm}^{+\res}(r) &= \left\{
     \begin{array}{llr}
       	0 															&\quad r < r_a	 				\\
       	L^B_{\elm}[\phi_{\elm}^-/W] 									&\quad r_a \le r < r_b 	\\
				L^B_{\elm}[\phi_{\elm}^-/W] + R^B_{\elm}[\phi_{\elm}^-/W]	&\quad r \ge r_b
     \end{array}
   \right. \nonumber\\
			&+ \Pi(x(r)) \int^r_{r_a} \frac{\tilde{\phi}_{\elm}^-}{W} S_{\elm}^I\, dr'	\c	
\end{align}
\begin{align}\label{eq:c_res_in}
	c_{\elm}^{-\res}(r) &= \left\{
     \begin{array}{llr}
       	0 															&\quad r > r_b	 				\\
       	R^B_{\elm}[\phi_{\elm}^+/W] 									&\quad r_b \ge r > r_a 	\\
				L^B_{\elm}[\phi_{\elm}^+/W] + R_{\elm}^B\left[\phi_{\elm}^+/W\right]	&\quad r \le r_a
     \end{array}
   \right. \nonumber\\
		&+ \Pi(x(r)) \int^{r_b}_r \frac{\tilde{\phi}_{\elm}^+}{W} S_{\elm}^I\, dr'\c
\end{align}
where the functionals $L^B_{\elm}$ and $R^B_{\elm}$ are given by
\begin{align}\label{eq:LB_RB}
	L^B_{\elm}[g(r)] 	&= \int^{r_a^+}_{r_a^-} g(r) S_{\elm}^B\, dr \nonumber\\
									 	&= \alpha_{\elm}(r_a) g(r_a) + \beta_{\elm}(r_a) g'(r_a)\c \\
	R^B_{\elm}[g(r)] 	&= \int^{r_b^+}_{r_b^-} g(r) S_{\elm}^B\, dr 			\nonumber\\
										&= -\alpha_{\elm}(r_b) g(r_b) - \beta_{\elm}(r_b) g'(r_b)\c
\end{align}
with
\begin{align}
	\alpha_{\elm}(x) &= -\frac{2(x-M)}{x(x-2M)} \phi^P_{\elm}(x) - \diff{\phi^P_{\elm}}{r}(x)		\c	\label{eq:alpha}\\
	\beta_{\elm}(x) 	&= \phi^P_{\elm}(x)\p			\label{eq:beta}
\end{align}
Splitting things in this way yields an interesting insight into the effective-source method. By
integrating the $\delta$-function terms analytically, we find that the scaling coefficients are
equivalent to worldtube jumps. The window function scheme of Vega and Detweiler is in fact
equivalent to Barack and Golbourn's worldtube method when one makes the particular choice of
window function given in Eq.~\eqref{eq:window_func}.

Outside the source region the weighting coefficients are constants given by
\begin{align}
	c_{\elm}^{-\res}(r) &= \left\{\begin{array}{ll}c^-_{\elm0} & r \le r_a\\ 0 & r > r_b  \end{array}\right.	\c	\\
	c_{\elm}^{+\res}(r)	&= \left\{\begin{array}{ll}0 &r < r_a \\ c^+_{\elm0} & r \ge  r_b \end{array}\right.	\p
\end{align}
We thus see that outside the source region the residual field and the retarded field coincide.

The source we have used in this work is sufficiently regular to allow for the calculation of the radial self-force by simply taking the derivative of the residual field at the particle. Differentiating Eq.~\eqref{eq:phi_inhom_variation_res} and using the definition of $c_{\elm}^{\pm\res}$ from Eqs.~\eqref{eq:cres+} and \eqref{eq:cres-} gives this derivative as
\begin{align}\label{eq:phi_res_deriv}
	\phi_{\elm}^{\res'}(r) = c_{\elm}^{+\res}(r) \tilde{\phi}_{\elm}^{+'}(r) + c_{\elm}^{-\res}(r)\tilde{\phi}_{\elm}^{-'}(r)\c
\end{align}
where the terms proportional to $c_{\elm}^{\pm\res'}(r)$ have cancelled.

\section{Numerical Implementation and Results}\label{sec:results}

The scalar-field self-force for a particle moving on a circular orbit about a Schwarzschild black hole was first calculated by Burko \cite{Burko-circular}. Explicit numerical results can be found in the work of Diaz-Rivera \etal~\cite{Diaz-Rivera}. In both works the retarded field was calculated numerically and regularized using the mode-sum scheme. In this work our goal is to demonstrate that our approach gives the same results. Our numerical scheme is described below; afterwards we present some sample results.

\subsection{Numerical algorithm}

The following steps describe how we compute the residual and retarded fields in practice.
\begin{enumerate}
	\item{(For $m\neq0$) Construct the boundary conditions at $r=r_\out$ and $r=r_\inn$ as described in Appendix~\ref{apdx:BCs}. }
	\item{(For $m\neq0$) Using standard ODE integration routines, numerically solve the homogeneous radial equation \eqref{eq:radial_eq} for $\tilde{\phi}_{\elm}^\pm$ between the boundaries and $r=r_a$. Let us denote the value of the field and its radial derivative at $r=r_a$ by $\tilde{\phi}^\pm_{\elm a}$ and $\tilde{\phi}^\pm_{\elm a}{}'$ respectively.}
	\item{(For $m\neq0$) Numerically solve the following coupled set of ordinary differential equations
\begin{align} 
	\square_{\elm}\tilde{\phi}_{\elm}^\pm &= 0\c \\ \diff{c_{\elm}^\pm}{r} &= \frac{\tilde{\phi}_{\elm}^\pm}{W}S^I_{\elm}\c
\end{align} 
from $r=r_a$ to $r=r_b$ with boundary conditions $\tilde{\phi}_{\elm}^\pm(r_a) = \tilde{\phi}^\pm_{\elm a}, \tilde{\phi}_{\elm}^\pm{}'(r_a) = \tilde{\phi}^\pm_{\elm a}{}', c^\pm_{\elm}(r_a)=0$. Let $c^\pm_{\elm b} \equiv c^\pm_{\elm}(r_b)$, $L^{B\pm}_{\elm} \equiv L^B_{\elm}(\tilde{\phi}_{\elm}^\pm/W)$ and $R^{B\pm}_{\elm} \equiv R^B_{\elm}(\tilde{\phi}_{\elm}^\pm/W)$. The residual field weighting coefficients are then given by
\begin{align}
	c_{\elm}^{+\res}(r) &= \left\{\begin{array}{ll}0 &r<r_a, \\ L^{B+}_{\elm} + c^+_{\elm}(r) &r_a \le r < r_b \\ L^{B+}_{\elm} + R^{B+}_{\elm} + c_{\elm b}^+ & r_b \ge r \end{array}\right.\c 	\\
	c_{\elm}^{-\res}(r) &= \left\{\begin{array}{ll}0 &r>r_b, \\ c_{\elm b}^- + R^{B-}_{\elm} - c_{\elm}^-(r) & r_b \ge r > r_a \\ L^{B-}_{\elm} + R^{B-}_{\elm} + c_{\elm b}^- & r<r_a \end{array}\right.\c 	
\end{align}}
\item{(For $m=0$) The homogeneous radial fields are given analytically by Eqs.~\eqref{eq:m=0_out} and \eqref{eq:m=0_in}. The residual field weighting coefficients are directly computed using Eqs.~\eqref{eq:c_res_out} and \eqref{eq:c_res_in}.}
\item{The residual radial field, $\phi_{\elm}^\res$, is then constructed via Eq.~\eqref{eq:phi_inhom_variation_res}. If the radial derivative is required Eq.~\eqref{eq:phi_res_deriv} is used. The full residual field is constructed using Eq.~\eqref{eq:Phi_decomp} with the replacement $\phi_{\elm}(r)\rightarrow\phi_{\elm}^\res(r)$. As our scalar field is real the $-m$ contributions to the sum are simply the complex conjugate of the $+m$ solutions.}
	\item{For comparison, the retarded radial field, $\phi_{\elm}^\ret$, is constructed via Eq.~\eqref{eq:phi_ret} and the individual $\ell$-modes of Eq.~\eqref{eq:Phi_decomp} (with the replacement $\phi_{\elm}(r)\rightarrow\phi_{\elm}^\ret(r)$) are regularized using Eqs.~\eqref{eq:field_mode-sum-formula} and \eqref{eq:force_mode-sum-formula}. }
\end{enumerate}

\subsection{Results}

Using the above algorithm we can compute the residual field at $r=r_0$. We find the result of this calculation agrees with the value of the regularized field, computed via the standard mode-sum approach, to a relative accuracy of better than $10^{-11}$. Furthermore, the puncture we have provided in this work is sufficiently regular to allow for the calculation of the radial self-force. By taking the radial derivative of the residual field at $r=r_0$ we find our method computes the radial self-force to a relative accuracy of better than $10^{-9}$. In Table \ref{table:results} we present some sample results. 

A key consideration in our calculation is the placement of the puncture boundaries $r_a$ and $r_b$. Formally, so long as $2M < r_a < r_0 < r_b$ the precise location of the boundaries will not effect the computed value of the self-force. In our practical numerical implementation though we find it best to arrange the boundaries such that $|r_0 - r_{\{a,b\}}| < 3M$ and $r_a > 3M$. Placing the boundaries within these constraints we find that the error in the computed self-force remains essentially constant (and in line with the results presented in Table \ref{table:results}). The reason we find it necessary to place these constraints is that the magnitude of the effective-source grows rapidly beyond this region and this causes difficulties for our numerical routines. If a wider source region is required then there are two immediate strategies that present themselves. Firstly, a higher-order puncture could be used in constructing the effective-source. This would act to smooth out the effective-source and could widen the region where it is small in magnitude. Secondly, an alternative window function would be employed that would attenuate the magnitude of the source away from the particle. If a formally compact source is not required then a simple Gaussian window-function might be suitable. Otherwise a compact smoothly attenuating source, such as that given in Ref.~\cite{Vega_etal:SF_in_3+1}, could be used. In this work we employed a Heaviside $\Pi$ function in order to make clear the connection between the window function and worldtube approaches. In other numerical schemes there may be more appropriate choices of window function. 

For both the effective-source and mode-sum calculations the higher-order regularization parameters of Ref.~\cite{Heffernan-Ottewill-Wardell} were used to accelerate the convergence of $\el$-mode sum. We envisage such an approach could be taken with a second-order-in-the-mass-ratio gravitational calculation; once the effective-source method has rendered the $\el$-modes of the fields finite, regularization parameters could be derived to increase the convergence rate of the $\el$-mode sum.

\begin{table}
\centering
\begin{tabularx}{\columnwidth}{r c c c c}
			\toprule
			& $r_0/M$			& eff.~source$\times10^{3}$ & mode-sum$\times10^{3}$ & rel.~diff.		\\
			\hline
$\Phi_0^\res$  & $6$		&	$5.454828078581$	&	$5.454828078597$	&	$3\times10^{-12}$	\\
$\partial_r \Phi_0^\res$	 & $6$		&	$0.16772830795$	&	$0.16772830804$	&	$5\times10^{-10}$	\\
\hline
$\Phi_0^\res$  & $10$		&	$-1.049793165979$	&	$-1.049793165983$	&	$4\times10^{-12}$	\\
$\partial_r \Phi_0^\res$	 & $10$		&	$0.013784482250$	&	$0.013784482234$	&	$2\times10^{-09}$	\\
\toprule
\end{tabularx}
\caption{
Sample results at $r_0=6M$ and $r_0=10M$. In both cases thirty $\el$-modes were computed and the effective-source boundaries where placed at $r_0\pm2M$. The third column gives the results of a C-code using the effective-source scheme presented in this work. The data in the fourth column was obtained using the same code to compute the retarded field and regularizing using the standard mode-sum procedure. The fifth column shows the relative difference between the results of the two calculations. For both orbits our results agree with those of Diaz-Rivera \etal~\cite{Diaz-Rivera}. Note that the data in this table has been adimensionalized (i.e., $\Phi^\res_0$ here $\equiv M/q\times\Phi^\res_0$).}\label{table:results}
\end{table}

\begin{figure}
	\center
\hspace{-0.2cm}	\includegraphics[width=8.7cm]{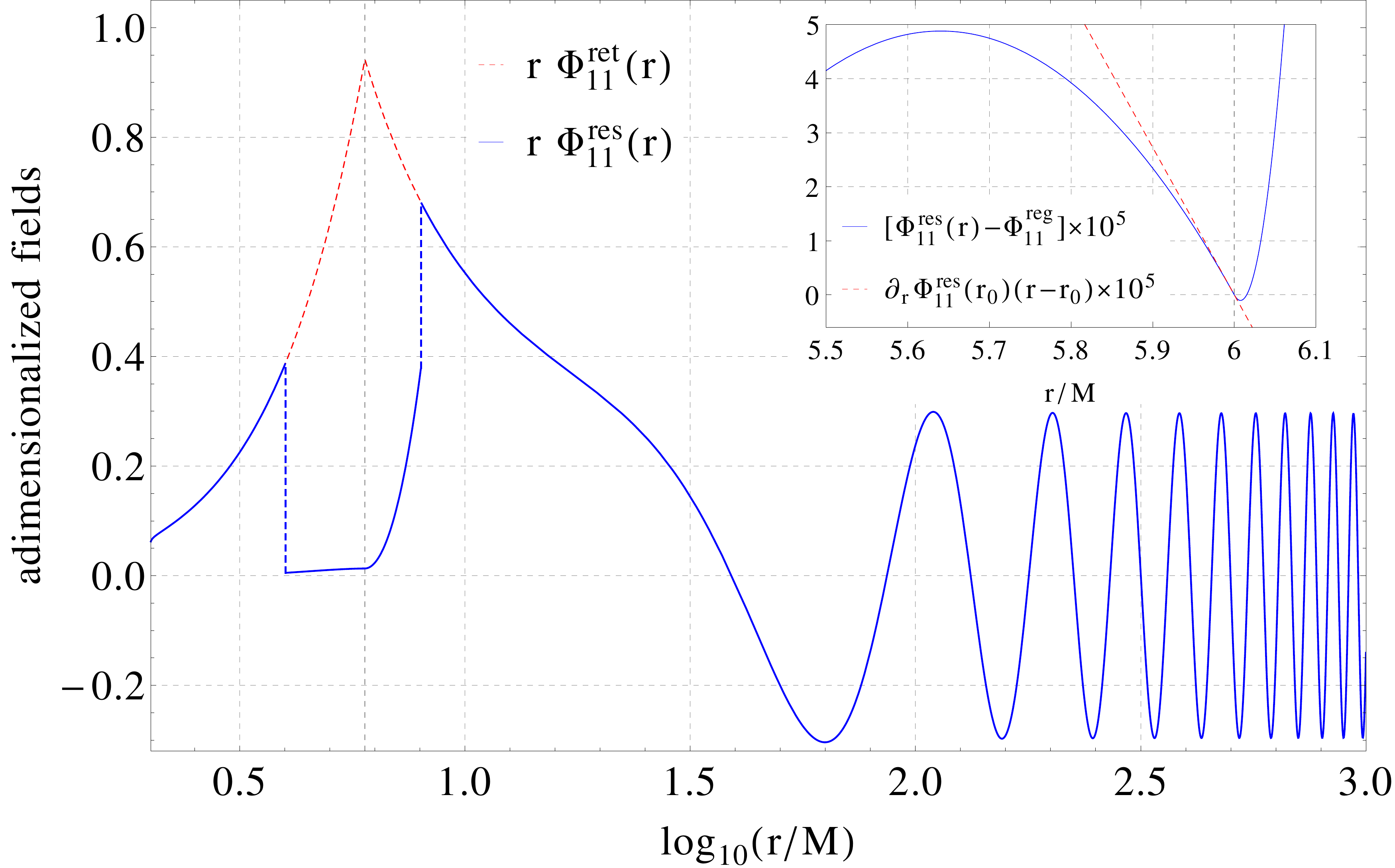}
	\caption{The $(\el,m)=(1,1)$ retarded and residual fields and the regularized field value for a particle at $r_0=6M$. The puncture boundaries are placed at $r_a=4M$ and $r_b=8M$. The dashed (red) curve shows the retarded field. Overlaid, the solid (blue) curve shows the residual field. Outside the puncture region the retarded and residual fields coincide. At $r=r_0$ the residual field takes the value of $\Phi^\text{reg}_{11}\equiv\Phi^\ret_{11}-B_\text{field} \approx 2.17628\times10^{-3}$. The inset plot shows the difference between the residual and regularized field value (solid, blue curve). Also shown in the inset is the derivative of the residual field at $r=r_0$ (dashed, red line). The slope of this line ($\approx -2.72955\times10^{-4} $) gives the $\el=1$ contribution to the radial self-force (there is no contribution from the $(\el,m)=(1,0)$ mode as only the $\el+m=\text{even}$ modes are non-zero). }\label{fig:l1m1}
\end{figure}

In Fig.~\ref{fig:l1m1} we plot the retarded, residual and regularized field value for the $(\el,m)=(1,1)$ mode for a sample orbit. The inset plot demonstrates the agreement between the regularized field calculated using the mode-sum and effective-source approaches.

\section{Relation between the puncture scheme and the mode-sum method: the ``zero-width worldtube'' approach}\label{sec:mode_sum_and_effective_source}

We now show that by carefully taking the limits $r_a\rightarrow r_0$ and $r_b\rightarrow r_0$ the usual
mode-sum regularization procedure can be recovered from the above effective-source scheme.
We also develop a ``zero-width worldtube'' approach in which we use alternative weighting
coefficients to rescale the homogeneous fields to give the regular field on the worldline, without
first having to construct the retarded field. In the proceeding discussion we find it useful
to make the following definitions:
\begin{align}
	c_{\elm0}^{+R} &\equiv L^B_{\elm}\left[\frac{\tilde{\phi}_{\elm}^-}{W}\right]_{r_a = r_0^-}\,,\quad c_{\elm0}^{-R} \equiv R^B_{\elm}\left[\frac{\tilde{\phi}_{\elm}^+}{W}\right]_{r_b=r_0^+}\c		\nonumber \\
	c_{\elm0}^{+S} &\equiv R^B_{\elm}\left[\frac{\tilde{\phi}_{\elm}^-}{W}\right]_{r_a = r_0^\pm}\,, \quad c_{\elm0}^{-S} \equiv L^B_{\elm}\left[\frac{\tilde{\phi_{\elm}}^+}{W}\right]_{r_b=r_0^\pm}\p\label{eq:RS_weighting_coeffs}
\end{align}
The $+$ and $-$ superscripts on $r_0$ indicate that the attached quantity should be evaluated by taking the
appropriate directed limit to the worldline. The $r_0^\pm$ cases can be evaluated in
either direction so long as the direction is taken consistently for both $c_{\elm0}^{+S}$ and $c_{\elm0}^{-S}$.

By comparing Eqs.~\eqref{eq:phi_inhom_matching} and \eqref{eq:phi_inhom_variation} and considering
Eqs.~\eqref{eq:c_res_out} and \eqref{eq:c_res_in} in the limit of $\{r_a,r_b\}\rightarrow r_0$ we can write the retarded field
scaling coefficients, $c^\pm_{\elm0}$, in terms of the puncture boundary functionals in the following way
\begin{align}
\label{eq:c0_again}
	c_{\elm0}^\pm &= c_{\elm0}^{\pm R} + c_{\elm0}^{\pm S} \nonumber \\
					&= \left[\alpha_{\elm}(r_0^-) - \alpha_{\elm}(r_0^+)\right]\frac{\tilde{\phi}_{\elm0}^\mp}{W_0}	 \nonumber\\
					&\,\,\,\,\,\,\,\,+ \left[\beta_{\elm}(r_0^-) - \beta_{\elm}(r_0^+) \right]\left( \frac{\tilde{\phi}_{\elm}^\mp}{W} \right)_{r=r_0}' \nonumber\\
					&= \left[\phi'^P_{\elm}(r_0^+) - \phi'^P_{\elm}(r_0^-)\right]\frac{\tilde{\phi}_{\elm0}^\mp}{W_0} \nonumber\\
					&= \kappa_{\elm} \frac{\tilde{\phi}_{\elm0}^\mp}{W_0} \c
\end{align}
where the third equality is seen to follow from Eqs.~\eqref{eq:alpha} and
\eqref{eq:beta} and the fourth equality is obtained by explicitly substituting for $\phi'^P_{\ell m}$.
The retarded field at the particle is then given, as before, by
\begin{align}
	\phi^\text{ret}_{\elm0} &= c_{\elm0}^+ \tilde{\phi}_{\elm0}^+ = c_{\elm0}^{-} \tilde{\phi}_{\elm0}^-	\nonumber \\
													&= \kappa_{\elm}\frac{\tilde{\phi}^+_{\elm0}\tilde{\phi}^-_{\elm0}}{W_0}	\p	\label{eq:phi_ret_2}
\end{align}

A more interesting use of the weighting coefficients defined above is to construct the regular
field at the location of the particle. By examining Eqs.~\eqref{eq:c_res_out} and \eqref{eq:c_res_in} it can be seen that
the regular field at the particle, $\phi^R_{\elm0}$, is given by
\begin{equation}
	\phi^R_{\elm0} = c_{\elm0}^{+R} \tilde{\phi}_{\elm0}^+ + c_{\elm0}^{-R} \tilde{\phi}_{\elm0}^-\p			\label{eq:phi_R}
\end{equation}
We call this method for calculating the regular field on the worldline the ``zero-width worldtube''
approach. We now show that this is identical to the usual mode-sum method and highlight an interesting property of the radial puncture. Using Eqs.~\eqref{eq:RS_weighting_coeffs} and \eqref{eq:LB_RB}, $\phi^R_{\elm0}$ can be written as
\begin{align}
	\phi^R_{\elm0} 	&= \left[\alpha_{\elm}(r_0^-) - \alpha_{\elm}(r_0^+)\right] \frac{\tilde{\phi}^+_{\elm0}\tilde{\phi}^-_{\elm0}}{W_0} - \beta_{\elm}(r_0)		\nonumber \\
						&= \kappa_{\elm}\frac{\tilde{\phi}^+_{\elm0}\tilde{\phi}^-_{\elm0}}{W_0} - \phi_{\elm}^P			\nonumber \\
						&= \phi_{\elm0}^\ret - \phi_{\elm0}^P.
\end{align}
We thus see that in the limit $\Delta r \rightarrow 0$ the radial puncture acts as an $\elm$-mode regularization parameter for the radial field. This leads to an $\elm$-mode regularization formula for the full field:
\begin{align}
	\Phi^R(x_p) = q \sum_{\el=0}^\infty\sum_{m=-\el}^\el \left(\phi^\ret_{\elm0} - \phi^P_{\elm0}\right) Y_{\elm}(\pi/2,\varphi_p)e^{-i\omega t_p}\p
\end{align}
The standard mode-sum regularization parameter can be derived from the above by summing over $m$.
Since the $\ell$ modes are invariant under rotations, this is equivalent to the standard procedure
for computing regularization parameters, where one sums over $m'$ in the rotated coordinate frame.
Explicitly this gives
\begin{align}
	B_\text{field} 	&= \sum_{m=-\el}^{\el} \phi^P_{\elm0} Y_{\elm} (\pi/2,0)	\nonumber \\
									&= \phi^P_{\el, m'=0}(r_0) Y_{\el,m'=0} (0, 0)\p
\end{align}
Similarly, the regularization parameters for the self-force are given by
\begin{align}
	(2\ell+1) A_r + B_r &= \sum_{m=-\el}^{\el} \partial_r \phi^P_{\elm0} Y_{\elm}(\pi/2,0)				\nonumber \\
		&= \partial_r \phi^P_{\el,m'=0} \big|_{r = r_0}  Y_{\el,m'=0} (0, 0).
\end{align}

For completeness, the $\elm$-mode contribution to the singular field (i.e., the $\elm$-mode regularization parameter given above) can be computed using the weighting coefficients in Eqs.~\eqref{eq:RS_weighting_coeffs} via
\begin{equation}
	\phi^S_{\elm0} = \phi^\text{ret}_{\elm0} - \phi^R_{\elm0}  =  c_{\elm0}^{+S} \tilde{\phi}_{\elm0}^+ + c_{\elm0}^{-S} \tilde{\phi}_0^-\label{eq:phi_S}\c
\end{equation}
Explicitly computing this quantity, as with $c^\pm_{\elm0}$ and $\phi_{\elm0}^R$ above, gives
\begin{align}
	\phi^S_{\elm0} 	= \beta_{\elm}(r_0) = \phi_{\elm0}^P\p
\end{align}

\section{Concluding remarks}\label{sec:concluding_remarks}

In this paper we have formulated the effective-source self-force approach in the frequency domain
and shown how it can be used to compute the self-force in the case of a scalar charge on a circular
orbit of a Schwarzschild black hole. The approach allows us to implement regularization of
individual $\ell,m$ modes of the retarded field including an arbitrary number of derivatives. We
validated our results against those obtained with mode-sum calculations on a mode-by-mode basis,
with differences which can be attributed to machine round-off. An obvious next step is to apply the
method to the gravitational case of the self-force on a point mass. In that case, the method
proceeds in essentially the same way, albeit with additional complexity from the tensor nature of
the field --- the full details of this calculation will be presented in a subsequent paper
\cite{Wardell-Warburton-Ottewill}.

The true advantage of our new method is not apparent in a first order calculation; after all, the
standard mode-sum scheme is perfectly adequate for a first order self-force calculation and there
is little benefit to the use of an effective-source approach. However, with the method
well-developed, its true use is its ability to handle situations where a mode decomposition alone
is insufficient to render the fields finite. For example, in gravitational self-force calculations
at second perturbative order \cite{Pound:nonlinear_GSF} there are terms which involve
$\rho_0^{-2}$, which will result in the spherical-harmonic modes of the field diverging as $\log
|\Delta r|$. Indeed, the field equation for the retarded field is not even well defined for $\Delta
r \to 0$, and an effective-source approach is essential.

For the example given here, we chose to use a second-order puncture as an approximation to the
singular field. With higher-order punctures being readily available, it may seem logical to
incorporate these higher orders into the effective source. However, it turns out that
doing so is neither necessary nor beneficial. Since we have shown here the equivalence of the
mode-sum and effective-source schemes, a much more straightforward, and computationally efficient
approach is to stick with a second order puncture and achieve improvements in accuracy by
subtracting the standard higher-order regularization parameters as a post-processing step. This
retains the accuracy benefits while keeping the computational cost of evaluating the effective
source to a minimum.

Although the focus of this paper has been on implementing the effective-source scheme in the
frequency domain, the methods developed may also be of use in improving the accuracy of time
domain calculations. Existing time domain approaches have relied on either numerical evolutions in
$2+1$ or $3+1$ dimensions, or on $1+1$ evolutions with a source which has been decomposed into
spherical-harmonic modes through numerical integrations \cite{Vega-Detweiler}. This severely limits the efficiency of any
code as the potentially-complicated effective source must always be evaluated on a three
dimensional grid. The punctures developed for our frequency-domain scheme present an ideal solution
to the problem; by decomposing \emph{analytically} into spherical-harmonic modes the computation
of the effective source is made more efficient by orders of magnitude, enabling the use of
accurate $1+1$ time domain calculations without the previous limitation of having a complicated and
computationally expensive effective source to evaluate.

Finally, we have focused here only on the relatively straightforward case of a non-spinning
Schwarzschild black hole. One would eventually want to apply the method to the more
astrophysically relevant case of a spinning Kerr black hole. The missing piece in that case would
be the adaption of the methods described here to the use of a \emph{spheroidal} harmonic basis.
While many of the methods would carry over unchanged, it is likely that the decomposition of the
puncture would be more involved.

\section*{Acknowledgements}
The authors thank Leor Barack, Adrian Ottewill, Adam Pound and Michael Boyle for helpful discussions. We also thank Sarp Akcay, Chris Kavanagh and Jeremy Miller for feedback on a draft of this work. N.W.'s work was supported by the Irish Research Council, which is funded under the National Development Plan for Ireland.
B.W. gratefully acknowledges support from Science Foundation Ireland under Grant No.~10/RFP/PHY2847
and from the John Templeton Foundation New Frontiers Program under Grant No.~37426 (University of
Chicago) - FP050136-B (Cornell University).
The authors additionally wish to acknowledge the SFI/HEA Irish Centre for High-End Computing
(ICHEC) for the provision of computational facilities and support (project ndast005b).

\appendix

\section{Boundary conditions for the scalar field}\label{apdx:BCs}

In our numerical calculation the radial domain extends from $r=r_\inn$ to $r=r_\out$ (how these are chosen in practice we will discuss below). In order to place boundary conditions at these finite radii we expand the asymptotic boundary conditions, given in Eq.~\eqref{eq:BCs_asymptotic}, in the form
\begin{align}
	\phi^+_{\elm} 		&= \frac{e^{i\omega_m r^\out_*}}{r} \sum_{k=0}^{k^+_{\max}} a_k^{+\elm} (\omega_m r_\out)^{-k}	\c\label{eq:BCs_out_expansion}	\\
	\phi^-_{\elm}		&= \frac{e^{-i\omega_m r^\inn_*}}{r} \sum_{k=0}^{k^-_{\max}} a_k^{-\elm} (r_\inn/M-2)^k					\c\label{eq:BCs_in_expansion}
\end{align}
where $r^\out_* = r_*(r_\out), r^\inn_* = r_*(r_\inn)$ and $k^\pm_{\max}$ are truncation indices.

Substituting these expansions into the radial equation \eqref{eq:radial_eq} gives recursion relations for the series coefficients $a_k^\pm$:
\begin{align}
	a_k^{+\elm} &= \frac{i}{2 k} \left[ (k(1-k) + \el(\el+1))a^+_{k-1} \right.			\nonumber\\
				& \left. + 2M\omega_m(k-1)^2 a^+_{k-2}\right]\c				
\end{align}
\begin{align}
	a_k^{-\elm} &= \frac{-1}{2 k(k-4iM\omega_m)}\left[((2k-1)(k-2) - \el(\el+1)\right.				\nonumber\\
				& - 12iM\omega_m(k-1))a^-_{k-1} +((k-2)(k-3)/2 		\nonumber\\
				& \left.- \el(\el+1)/2 - 6iM\omega_m(k-2))a^-_{k-2} \right.		\nonumber\\
				& \left.- iM\omega(k-3)a^-_{k-3}\right]	\p
\end{align}	
In constructing the homogeneous solutions to the radial equation we set $a_0^\pm=1$ after which all the other terms in the series are uniquely determined by taking $a_{k<0}^\pm = 0$. The truncation indices are chosen so that the relative contribution of the $k=k^\pm_{\max}$ term, with respect to the first term, is below $10^{-14}$.

The extent of the radial domain is chosen so that $\omega r_\out \gg 1$ and $r_\inn/M-2 \ll 1$ which ensures the series \eqref{eq:BCs_out_expansion} and \eqref{eq:BCs_in_expansion} converge rapidly. In practice we find that setting $r_\inn=2.001M$ and $r_\out=10/\omega$ suffices.

\bibliographystyle{apsrev4-1}

\end{document}